# Advancements in scientific data searching, sharing and retrieval

Ranjeet Devarakonda[1], Giri Palanisamy[1]

[1] Oak Ridge National Laboratory PO Box 2008 MS 6407, Oak Ridge, TN 37831 USA

OCT 2010

In the recent years, there has been significant advancement in the areas of scientific data management and retrieval techniques, especially in terms of standards and protocols for archiving data. Oak Ridge National Laboratory Distributed Data Archive Center for biogeochemical dynamics is making efforts in building advanced toolsets for these purposes. Mercury is a web-based metadata harvesting, data discovery and access system, built for researchers to search for, share and obtain biogeochemical data. Originally developed for single National Aeronautics and Space Administration (NASA) project, Mercury now used over fourteen different projects across three US federal agencies.  Mercury renders various capabilities including metadata management, indexing, searching, data sharing, and also software reusability.

Mercury system harvests the structured data from several data provider servers around the world. The harvested files are indexed against Solr search API consistently, so that it can render various search capabilities such as simple, fielded, spatial and temporal searches across a span of projects ranging land, atmosphere, and ocean ecology.  Mercury also provides implementation techniques for data sharing between data providers using OAI-PMH. This chapter will talk about data harvesting of structured metadata, efficient ways of indexing and searching using Solr search API.

A key conclusion in a recent United States Interagency Working Group on Digital Data (IAWGDD) report on harnessing the power of digital data for science and society is the role of communities of practice in effective quality control, preservation, distribution, interpretation, and use of digital data (NSTC, 2009).  A given researcher may, however, participate in multiple communities of practice, and may also need to draw on data from communities outside those he or she normally participates in.  The data generated by a researcher, or by any other data generator, is potentially of use to multiple communities of practice and scientific disciplines.  While that data may be archived in a particular repository serving one or more particular communities, that data may need to be discoverable and usable by multiple communities.

General-purpose search engines, such as Google and Bing, are currently generally ineffective at discovering scientific data, in part because of the specific semantics associated with a particular search and because those search engines generally perform full text, rather than fielded searches of structured documents.  A Google search on the term "Eagles" lacks the context to distinguish between multiple different meanings, whereas a data repository serving a biology community can presume that the searcher is referring to members of the genus Haliaeetus.

Advances in the practice of scientific data management, the tools for managing data, the standards for data formats and metadata formats, and the understanding of the value of digital data have created a wide range of digital repositories focused on different applications.  Nor are these repositories necessarily distinct.  There may be a number of different repositories serving field ecologists, with distinctions based on funding agency, country, organizational affiliation, or other artifacts of historical origin.  These repositories generally have search tools that work within their particular holdings, but are often unable to search across the holdings of other repositories, due to various technical and sociological factors.  From the end user perspective, this situation is problematic, as a comprehensive search for available digital data relevant to a research topic is nearly impossible, requiring knowledge of multiple repositories and the particular search interfaces of those repositories.

Multiple approaches have been used for enabling search across multiple repositories, such as the Z39.50 (Information Retrieval Standard, 1997) distributed search method.  Distributed searches, however, can be problematic, both for response time and uptime.  Search results can only be presented to the user as quickly as the slowest search agent returns (plus some processing time if the results are to be integrated) and the composite uptime is the product of the individual uptimes.

As a result of the problems with distributed searches, repositories have turned to a harvest and index approach as a means to ensure rapid response, enable full integration of metadata from multiple sources, and provide acceptable uptime.  However, harvesting can be an inefficient process, particularly if the metadata are completely reharvested regularly as a means to ensure that source changes are propagated into the search results.

Mercury (Devarakonda et al., 2010) is an open-source toolset for metadata authoring, harvesting, indexing, and searching which implements a variety of harvesting protocols and provides a coherent view of metadata across a range of metadata standards, including Federal Geographic Data Committee Content Standard for Digital Geospatial Metadata (FGDC CSDGM), Ecological Markup Language (EML), Global Change Master Directory's Directory Interchange Format (GCMD DIF), Dublin Core and ISO 19115. Mercury's architecture includes 1) a harvesting engine to collect various metadata records from publically available folders, web sites, ftp sites, and other network accessible locations; 2) a powerful indexing engine based on Apache Lucene and SOLR that can index billions of records; and 3) a service oriented architecture based search engine, which can perform searches and distribute results through web user interfaces, web services, RSS feed, and portlets.